\documentclass[12]{article}
\usepackage[dvips]{graphicx}
\def\bB{\mathbf{B}}
\def\bb{{\cal B}}
\def\bE{{\bf E}}
\def\be{{\bf e}}
\def\bJ{\mathbf{J}}
\def\Ce{{\cal C}}

\def\bx{{\bf x}}
\def\by{{\bf y}}
\def\x{{\bf e}_1}
\def\y{{\bf e}_2}
\def\z{{\bf e}_3}
\def\k{{\bf e}_k}
\def\s{{\bf e}_}
\def\e{e^}
\def\w{\wedge}

\def\C{\mathcal{C}}
\def\R{\mathcal{R}}
\def\bH{\mathcal{H}}

\def\tri{\triangle}
\def\no{\noindent}

\def\suk{\sum_{k=1}^3}

\def\no{\noindent}
\def\beq{\begin{equation}}
\def\eeq{\end{equation}}

\def\E^b(r){{\bf e_2}}

\def\ba{{\bf a}}
\def\bb{{\bf b}}
\def\bc{{\bf c}}

\def\bB{{\bf B}}
\def\bv{{\bf v}}

\def\bx{{\bf x}}
\def\by{{\bf y}}
\def\bv{{\bf v}}

\def\ba{{\bf a}}
\def\bp{{\bf p}}
\def\bv{{\bf v}}
\def\bb{{\bf b}}
\def\w{{\wedge}}

\def\bc{{\bf c}}

\def\R{I\!\!R}
\def\H{I\!\!H}
\def\C{I\!\!\!C}

\def\s{{\bf e}}
\def\no{\noindent}

\begin{document}
\title{Special relativity in complex vector algebra}
\author{Garret Sobczyk   \\
Departamento de Actuar\'ia y Matem\'aticas \\ Universidad de Las Am\'ericas - Puebla,\\ 72820 Cholula, Mexico}
\maketitle
\section{Introduction}
   Special relativity is one of the monumental achievements of physics of the 20th Century. 
 Whereas Einstein used a coordinate based approach \cite{E1905}, which obscures 
 important geometric aspects of this fundamental theory, many coordinate free geometric
  languages have since been developed.  In \cite{H74}, D.
 Hestenes showed how the ideas of special relativity can be elegantly expressed in
 {\it space-time algebra}. 
  The purpose of this paper is to examine the fundamental ideas of special relativity in a
 complex vector-based language that is the natural generalization of the 
 Gibbs-Heaviside vector algebra of 3-dimensional space \cite{S4}.   
\section{The algebra $\C_3$ of complex vectors.}
Let $\bb=\{\x,\y,\z\}$ be an orthonormal basis for a complex 3-dimensional vector space $\Ce^3$, taken together with 
a {\it complex scalar product} $A\circ B$, defined by 
\beq
\label{puntoc}
A\circ B :=\suk\alpha_k\beta_k=\alpha_1\beta_1+\alpha_2\beta_2+\alpha_3\beta_3= B\circ A
\eeq
for $A=\suk \alpha_k\k \in \Ce^3$ and $B=\suk \beta_k\k\in \Ce^3$ where $\alpha_j,\beta_k \in\Ce$. This means that
  \beq \x\circ \x=\y\circ \y = \z \circ \z = 1 \ {\rm and} \ 
    \s_k \circ \s_j =0       \label{othonorm} \eeq
for all $j,k=1,2,3,$ and $j\ne k$. 

Our immediate objective is to introduce more structure, together with a comprehensive geometric interpretation,
 and turn the $3$-dimensional complex vector space $\Ce^3$ into
a $4$-dimensional {\it complex vector algebra} $\Ce_3$ of space-time.

Analogous to vector analysis, we define 
a {\it complex vector product}
\beq A\otimes B:= i \det \pmatrix{ \x & \y & \z \cr
					   \alpha_1 & \alpha_2& \alpha_3 \cr
					   \beta_1 & \beta_2& \beta_3} = - B\otimes A,
\label{vectorproduct} \eeq
where $i\in \C$ is the imaginary unit with $i^2=-1$.

\begin{figure}
 \begin{center}
\includegraphics[scale=0.35]{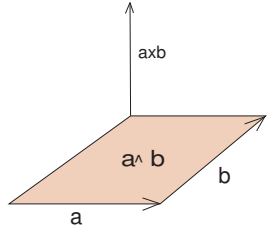}
\caption{\label{fig:epsart} The vector $\bf a$ is swept out along the vector $\bf b$ to form the
bivector ${\bf a} \w {\bf b}$. The vector $\ba \times \bb$ is the right-handed normal to this plane.}
\end{center}
\end{figure}  

For the real vectors ${\bf a}=\suk a_k \k$ and ${\bf b}=\suk b_k \k$ for $a_k,b_k\in \R$,
the usual dot and cross products are defined by
$$\mathbf{a\cdot b}:=\sum^3_{k=1}a_kb_k=a_1b_1+a_2b_2+a_3b_3$$ 
and
\[\mathbf{a\times b}:=\det\pmatrix{\x&\y&\z\cr
					   a_1&a_2&a_3\cr
					   b_1&b_2&b_3 } . \]
For the real vectors ${\bf a}, {\bf b}$, we see that	
$${\bf a}\circ {\bf b}={\bf a}\cdot {\bf b}$$
and
$$\mathbf{a \otimes b}=i({\bf a\times b}):={\bf a} \w {\bf b}.$$
We give $i({\bf a}\times {\bf b})$ the interpretation of
  the {\it bivector} ${\bf a}\w {\bf b}$, or {\it directed area segment}
having the right-handed normal vector ${\bf a} \times {\bf b}$. In the
sense of Grassmann, the bivector ${\bf a} \w {\bf b}$ is the directed area obtained by sweeping the vector $\bf a$ out
along the vector $\bf b$. See Figure 1.

Unlike the dot and cross products, which are real linear, the complex scalar and complex 
vector products are complex linear
\[ \alpha(A\circ B)=(\alpha A)\circ B=A\circ(\alpha B)\ \ {\rm and} \ \ 
\alpha(A\otimes B)=(\alpha A)\otimes B=A\otimes(\alpha B) \]
 for all $\alpha\in\C$.

For real vectors $\ba, \bb, \bc$, we have
\beq
\label{vol}
{\bf a\circ (b\otimes c)}=i[{\bf a\circ (b\times c)}]=i[{\bf a\cdot (b\times c)}]:={\bf a\wedge (b\wedge c)},
\eeq
suggesting that ${\bf (a\wedge b)\wedge c}={\bf a\wedge (b\wedge c)}$ be given 
the geometric interpretation of the {\it trivector}, or directed element of
volume, obtained by sweeping the bivector $\bf a \w b$ out along the vector $\bf c$.
Choosing $\bf a=\x, b=\bf \y, c=\z$ in the above, gives
$$i=\x\circ (\y\otimes\z)=\x\wedge\y\wedge\z ,$$
so $i$ has the geometric interpretation of a unit trivector or \emph{unit pseudoscalar}. See Figure 2.

\begin{figure}
\begin{center}
\includegraphics[width=7cm]{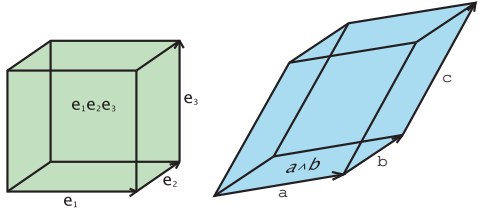}
\caption{\label{fig:epsart} The trivector ${\bf a}\w {\bf b} \w {\bf c}$ is formed by sweeping the
bivector ${\bf a}\w {\bf b}$ out along the vector $\bf c$. Also shown is the unit pseudoscalar $i$. }
\end{center}
\end{figure}  

Combining the complex scalar and complex vector products gives the associative {\it complex geometric product}
\beq \label{gp}
AB:=A\circ B + A\otimes B
\eeq
of the complex vectors $A$ and $B$. Because of the properties of the complex scalar and complex
vector products, (\ref{gp}) is equivalent to the pair of identities
\beq  A\circ B:=\frac{1}{2}(AB+BA) \ {\rm and} \ A\otimes B:=\frac{1}{2}(AB-BA) , \eeq
so the complex scalar and complex vector products could have equally well been defined in terms of
the more fundamental complex geometric product.

The {\it complex vector algebra} $\Ce_3$ of the complex vector space $\Ce^3$ is defined by
\beq 
\begin{array}{lll}
\Ce_3  =\C\oplus \Ce^3 =\{\alpha + A | \  \alpha \in \Ce, A\in \Ce^3  \},
\end{array}
\eeq
taken together with the complex geometric product (\ref{gp}). The elements of $\Ce_3$ form a closed complex $4$-dimensional
linear space with the basis $\{1,\x,\y,\z \}$, which is algebraically 
isomorphic to the Pauli algebra of complex $2\times 2$ matrices. The {\it space-time algebra} $\Ce_3$
 for special relativity is the natural generalization of the universally known
Gibbs-Heaviside vector algebra of space. 

\section{Special relativity in complex vector algebra.}

  All of the observables of space-time can be viewed as elements of the complex vector algebra. The observables
are time (real scalars), vectors, bivectors, and trivectors (pseudoscalars). The basic operations in space-time are
active and passive rotations and active and passive velocity transformations or boosts.  

An {\it active} rotation is defined by
\beq
\label{rotation}
{\bf x^\prime}=\e{-{1\over 2}\theta i\z}{\bf x}\e{{1\over 2}\theta i\z},
\eeq 
where the physical vector $\bf x$ is actively being rotated in the plane of the bivector $\be_{12}=i\z$ through
the Euclidean angle $\theta$. In contrast, for a {\it passive} rotation, the vector $\bx$ stays fixed, whereas
the reference frame of the observer is rotated in the opposite direction.

\begin{figure}
\begin{center}
\includegraphics[width=7cm]{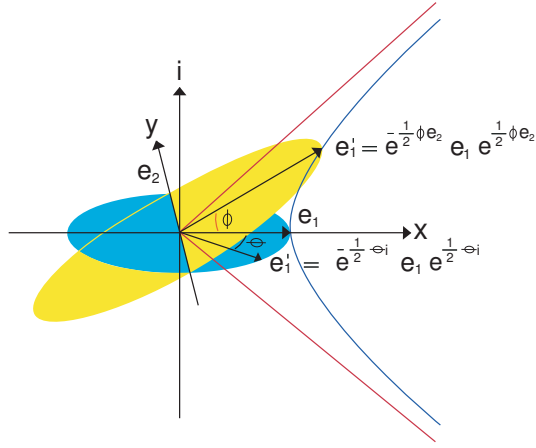}
\caption{\label{fig:epsart} An active Lorentz rotation and an active Lorentz boost are illustrated.}
\end{center}
\end{figure}

In complex vector algebra an {\it active} velocity transformation is given by
\beq
\label{boost}
{\bf x^\prime}=\e{-{1\over 2}\phi\y}{\bf x}\e{{1\over 2}\phi\y},  
\eeq
where the physical vector $\bx$ of a rest frame is being given a velocity in the direction of the vector $\y$ at the
speed $v/c:=\tanh \phi$ into the vector $\bf x^\prime$ (of a different observer), where $c$ is the speed of light. 
In contrast, for a {\it passive} boost the vector $\bx$ stays fixed, whereas
the reference frame of the observer is given a speed of $v/c=\tanh \phi$ in the opposite direction. Note,
for both a rotation (\ref{rotation}) and a boost (\ref{boost}), that ${\bx^\prime}^2=\bx^2$, so that the square of a
complex vector is preserved under both active rotations and boosts. 

Letting $\bf x=\x$, the calculation
\[ {\bf \x^\prime}=\e{-{1\over 2}\phi\y}{\bf \x }\e{{1\over 2}\phi\y}={\bf \x }\e{\phi\y} \] 
\[  =\x(\cosh \phi + \y \sinh \phi)=\x \cosh \phi + i \z \sinh \phi .   \]
shows that the active boost of $\x$, through the hyperbolic angle or {\it rapidity} $\phi$ in the direction
of the vector $\y$, gives $\x^\prime$, which is a linear combination of the vector $\x$ and
the bivector $i \z$. 
The complex vector algebra allows rotations and velocity transformations to be put on the same
footing with an equally immediate geometric interpretation. See Figure 3.

The concept of an active Lorentz boost is not used by physicists or mathematicians today, although
early attention was called to it in \cite{S4}. Perhaps the reason why active boosts have never been
recognized is that vectors and bivectors have never been
explicitly added together as elements of the same linear space, despite the fact that this is exactly what
Einstein's special theory of relativity calls for \cite{S1}. 

\subsection{Event Horizon of an inertial system.}

All events in the Universe take place in the space-time algebra $\Ce_3$.
The {\it event horizon} of an {\it inertial system} is a subset $\H\subset \Ce_3$:
$$\H=\{X|X=\sum_{\mu=0}^3 x_\mu \be_\mu \mbox{ where} \ \ x_\mu \in\Re\}$$
or
$$X=ct+\bx$$
where $x_0 =x_0 \be_0 =ct$ and $\bx =\suk x_k\k$. The {\it event} $X$ occurs at the {\it time} $t$
and at the {\it place} defined by the position vector $\bx$, as measured by an observer in the
inertial system $\H$, where $c$ is the speed of light. 
If $X(t)=ct+{\bf x}(t)$ is the {\it space-time} history of a particle 
in the inertial system $\H$, then the {\it space-time velocity } of the particle is 
  \beq V=\frac{dX}{dt}=c+\frac{d{\bf x}}{dt}=c+{\bf v}=c(1+\frac{\bv}{c}), \label{stvelocity} \eeq
where ${\bf v}=\frac{d\bf x}{dt}$ is the velocity of the particle at the time $t$. If
$X(t)$ is the space-time line of an inertial observer, then its space-time velocity 
$V(t)=\frac{dX}{dt}=c$, so
the velocity of a inertial observer in $\H$ is always ${\bf v}=0$. 

 The orthonormal rest frame
$\{\k \}$, for $k=1,2,3$, of a given inertial system $\H$ is characterized by the
algebraic properties
  \beq  \x^2=\y^2=\z^2=1, \be_j \be_k=-\be_k \be_j, \ {\rm and}
 \ \x\y\z=i \label{orthoframe} \eeq
where $j\ne k$.

Now let $\H^\prime$ be the event horizon of an inertial system moving along the $x$-axis with
the velocity $v \x$ as seen by an observer in $\H$. 
Then the event horizons are related by the {\it universal mapping}
  \beq X^\prime = X \exp( \phi \x), \label{horizon} \eeq
and the orthonormal rest frames by the boost
 $\be_k^\prime =e^{-\frac{1}{2}\phi \x} \be_k e^{\frac{1}{2}\phi \x} $
for $k=1,2,3$, where the hyperbolic angle $\phi$ satisfies
$\frac{v}{c}=\tanh \phi$. In Einstein's 1905 paper about 
special relativity \cite{E1905}, he discusses
the constant speed of light in all inertial systems, and gives an elaborate way of
measuring relative time in different inertial systems.
In this formulation of relativity, each observer in his or her event horizon 
measures position and time in the usual Newtonian way. All of the additional assumptions of special relativity that
go beyond the Newtonian-Galilean World are specified in (\ref{horizon}), which explicitly expresses
how a boost changes the way a given event is meassured in a different inertial system. 

When $v<<c$, the event horizons $\H$ and $\H^\prime$ are related by
  \[ X^\prime = ct^\prime +\bx^\prime = (ct + \bx)\e{\phi \x}\widetilde = (ct + \bx)(1+\frac{v}{c} \x) \]
  \[     = ct + \frac{vx}{c}+\bx +vt \x + \frac{v}{c} \bx \otimes \x \widetilde = ct+\bx + vt \x ,\]
leading to the so-called Galilean transformation of coordinates 
   \[  t^\prime = t \ {\rm and} \ \bx^\prime = \bx+vt \x , \]
 or equivalently,
 $$
\begin{array}{l}
t'=t\\
x'=x+vt\\
y'=y\\
z'=z.\\
\end{array}
$$
     
Suppose now that $X(t)=ct+{\bf x}=ct+x \x+y \y+z \z$ is the space-time history of a particle
moving in the inertial system $\H$, and that $X^\prime=ct^\prime+x^\prime \x^\prime+y^\prime \y^\prime+z^\prime \z^\prime$
is the space-time history of the same particle but as seen in the inertial sytem $\H^\prime$.
We now relate the coordinates $\{t,x,y,z \}$ as measured in $\H$ to the
coresponding coordinates $\{t^\prime,x^\prime,y^\prime,z^\prime \}$ as measured in $\H^\prime$. The relations
$ X^\prime = X \exp( \phi \x)$, and
 $\be_k^\prime =e^{-\frac{1}{2}\phi \x} \be_k e^{\frac{1}{2}\phi \x} $ for $k=1,2,3$ imply that
\beq e^{\frac{1}{2}\phi \x}X^\prime e^{-\frac{1}{2}\phi \x}=
e^{\frac{1}{2}\phi \x} X e^{\frac{1}{2} \phi \x}, \label{basic} \eeq
from which it follows that
  \[ ct^\prime + x^\prime \x^\prime = (ct+x\x ) e^{\phi \x}=(ct+x\x)(\cosh \phi +\x \sinh \phi)=
       \cosh \phi(ct+x\x)(1 +\x \frac{v}{c} )                    \]
and  
   \[  y^\prime \y +z^\prime \z =y \y +z \z, \]
so the corresponding Lorentz transformation of the coordinates are
$$
\begin{array}{l}
t'={t+{vx\over c^2}\over\sqrt{1-({v\over c})^2}}\\
x'={x+vt\over\sqrt{1-({v\over c})^2}}\\
y'=y\\
z'=z.\\
\end{array}
$$

From the basic relationship (\ref{basic}), and noting that $\x^\prime = \x$, we can identify the right-hand side of
this equation,
\beq
      \e{{1\over 2}\phi\x}X \e{{1\over 2}\phi\x},  \label{passive}
\eeq
 as representing a {\it passive} Lorentz boost in the direction of $\x$ with speed $v$, 
 whereas we have already seen in (\ref{boost}) that the left-hand side of this equation
 \beq \e{{1\over 2}\phi\x^\prime}X^\prime \e{-{1\over 2}\phi \x^\prime} 
   \label{active}  \eeq
represents an active Lorentz boost in the dirction of $-\x$ with speed $v$.
 
We see from (\ref{basic}), that an active boost of the event horizon $\H^\prime$ in the direction of the
$x^\prime$-axis at the speed $-v$ is equivalent to a passive boost of the event horizon $\H$ in the direction
of the $x$-axis at the speed $v$.

\subsection{The proper conjugation of an inertial system and space-time inversion}

We have seen in (\ref{orthoframe}) that every 
 {\it orthonormal basis} $\{\x,\y,\z\}$ of complex vectors of $\Ce^3$, satisfying the property
that
  \[ \x \y \z =i, \quad {\rm and} \quad \x^2=\y^2=\z^2=1, \]
defines the {\it rest frame} of an inertial system. What we need is a mechanism for
distinquishing between the vectors and bivectors of one inertial system from the vectors and bivectors
of any other inertial system. 

\bigskip

\no {\bf Definition} By a {\it conjugation} on the complex vector algebra $\C_3$, we mean an operator
$\overline{\cal A}$, defined for all ${\cal A},{\cal B}\in \C_3$, which satisfies
\begin{itemize} \label{conjugation}
\item[]{1.} $\overline{x+iy}=x-iy$ for all $x,y\in \R$,
\item[]{2.} $\overline{{\cal A}+{\cal B}}=\overline{\cal A}+\overline{\cal B}$,
\item[]{3.} $\overline{\cal{AB}}=\overline{\cal B}\ \overline{\cal A}$,
\item[]{4.} $\overline{\overline{\cal A}}={\cal A}$. 

\end{itemize} 

Let an inertial system $\H$ be given together with its orthonormal frame
$\{\x,\y,\z\}$ of vectors. The {\it proper conjugation} of $\H$ is defined in such a way that it takes
real vectors $\bx = x \x + y \y+z \z$ into themselves and changes the sign of imaginary vectors (bivectors).
More precisely, if $\bx, \by$ are real vectors of $\H$, then the proper conjugation of $\H$ satisfies
  \beq \overline \bx = \bx, \quad {\rm and} \quad \overline{\bx \by} =\overline \by \ \overline \bx=\by \bx.  
                        \label{properconjugation} \eeq
    It follows that an inertial system $\H$ defines a unique splitting of a complex vector $F$ into real and
  imaginary vector parts:
    \[ F= \frac{1}{2}(F + \overline F) +  \frac{1}{2}(F - \overline F)= \bE+ i \bB \]
  where $\bE $ and $\bB $ are real vectors of $\bH $, \cite{S2}.  

   Another important operation on $\C_3$ is {\it complex-vector inversion}. Given ${\cal A}=\alpha +A\in \C_3$,
the {\it inversion} of $\cal A$, defined by
   \beq {\cal A}^- := \alpha - A, \label{inversion} \eeq
has the effect of changing the sign of the complex vector part of ${\cal A}\in \C_3$. Complex-vector inversion
(\ref{inversion}) satisfies the last three properties of a conjugation operator 
given in the definition, but leaves unchanged
the complex scalar part of ${\cal A}$.

   Given an event $X=ct + \bx$ in an inertial system $\H$, the {\it space-time interval} of $X=ct+\bx$ is defined by
   \beq  |X|_{st}^2:=XX^-=c^2 t^2-\bx^2.  \label{st-interval}   \eeq
If $X^\prime = ct^\prime + \bx^\prime=X\e{\phi \x}$, then the easy calculation
  \beq  c^2{t^\prime}^2-{\bx^\prime}^2=X^\prime {X^\prime}^- =  \e{-\phi \x}XX^-\e{\phi \x}=XX^-
           =c^2 t^2-\bx^2 \label{crucial} \eeq
shows that the space-time interval $|X|_{st}^2$ of the same event 
as measured in a different inertial system is preserved. This is the
crucial result upon from which all of the surprising results of special relativity follow.

\section{Differentiation in space-time} 

  In standard vector analysis, the {\it gradient} of a scalar field, and the 
{\it divergence} and {\it curl} of a vector field are introduced in terms of the
nabla operator
  \beq \nabla_\bx = \x \frac{\partial}{\partial x} + \y \frac{\partial}{\partial y} 
 + \z \frac{\partial}{\partial z} , \label{nabla}  \eeq
for the vector variable $\bx=x \x +y \y + z \z$.

For space-time, we introduce the {\it space-time nabla operator} $\tri_X$,
   \beq   \tri_X := \frac{1}{c}\frac{\partial}{\partial t} + \nabla_\bx \label{stnabla} \eeq
for the space-time variable $X=ct+\bx$ of a given inertial system $\H$. From $\nabla_\bx \bx =3$,
it easily follows that $\tri_X X=1+3=4$, showing that the the variable $X$ has 4 degrees of freedom
in the space-time horizon $\H$. Multiplying both sides of the last equation on the right by $\e{-\phi \x}$,
 and on the left by $\e{\phi \x}$, gives
   \[ \e{-\phi \x}\tri_X X \e{\phi \x} =\e{-\phi \x}\tri_X X^\prime = \tri_{X^\prime} X^\prime =4, \]
or 
  \beq \tri_{X^\prime}=\e{-\phi \x}\tri_X , \label{chainrule} \eeq
which is easily shown to be a consequence of (\ref{horizon})
and the chain rule for the change of the variables $\{t,x,y,z\}$ to the variables $\{t^\prime,x^\prime,y^\prime,
z^\prime \}$.

 \subsection{The electromagnetic field}

    An {\it electromagnetic field} $F=F(X)$ observed in the event horizon $\bH$ has
the form
    \beq F=\bE +  i \bB \label{electromag} \eeq
where $\bE=\frac{1}{2}(F + \overline F) $ is the electric field part and 
     $\bB = \frac{1}{2}(F - \overline F)$ is the magnetic field part.
Alternatively, the same electromagnetic field $F=F(X^\prime)$, seen by an observer
in the spacetime horizon $\H^\prime$ is 
    \beq F=\bE^\prime +  i \bB^\prime \label{electromag} \eeq
where $\bE^\prime=\frac{1}{2}(F + \widetilde F) $ is the electric field part and 
     $\bB^\prime = \frac{1}{2}(F - \widetilde F)$ is the magnetic field part.
By writing $F=F(X)=F(X^\prime)$, we are expressing that the electromagnetic
field $F$ seen at $X$ in $\H$ is the same field as seen at $X^\prime \in \bH^\prime$.

  Of course, the electromagnetic field $F$ must satisfy {\it Maxwell's Equations},
  \beq  \tri_X F =( \frac{1}{c} \frac{\partial}{\partial t} +
 \nabla_\bx )(\bE+i \bB)=4 \pi (\rho - \frac{1}{c}{\bf J}), \label{maxwell}  \eeq
where $\rho$ is the {\it charge density} and $\bf J$ is the {\it current density} 
\cite[p.182]{Jack} at the event $X$ in
the inertial system $\bH$. Multiplying this last equation on the left-hand side by $\e{-\phi \x}$ gives
  \[  \tri_{X^\prime}F =\e{-\phi \x} \tri_X F =4 \pi \e{-\phi \x}  (\rho - \frac{1}{c}{\bf J})=
       4 \pi (\rho^\prime - \frac{1}{c}{\bf J^\prime}),  \]
which is Maxwell's equation in the inertial system $\H^\prime$. The 4 standard Maxwell equations can be
recovered by separating the equation (\ref{maxwell}) into its respective scalar, vector, bivector, and
pseudoscalar parts, giving
   \begin{itemize}
\item[] $$ \nabla_\bx \cdot \bE = 4 \pi \rho, \ \ \ \ \frac{1}{c} 
\frac{\partial} {\partial t} \bE - \nabla \times \bB= -\frac{4 \pi}{c} \bJ $$
\item[]  $$ \frac{1}{c} \frac{\partial} {\partial t} \bB + \nabla \times \bE= 0, \ \ \ \ \nabla_\bx \cdot \bB=0.$$
\end{itemize} 

  Maxwell's equations can also be formulated in terms of the {\it scalar potential} $\Phi$ and
the {\it vector potential} $\bf A$ \cite[p.179]{Jack}. Noting that 
  \[  \tri_X \tri_X^- {\bf A} = \frac{4 \pi}{c} \bJ \ \ {\rm and} \ \ \tri_X \tri_X^- \Phi = 4 \pi \rho  \]
where  $\tri_X \tri_X^-= \frac{1}{c^2}\frac{\partial^2}{\partial t^2} - \nabla_\bx^2$ is the
Laplacian, 
Maxwell's equation becomes
  \beq \tri_X \tri_X^- (\Phi - {\bf A})= 4\pi (\rho - \frac{\bJ}{c}). \eeq
In order to verify that this last equation is equivalent to (\ref{maxwell}), we must assume the
{\it Lorentz condition} that $\frac{1}{c} \frac{\partial \Phi}{\partial t} + \nabla_\bx \cdot {\bf A}=0$,
\cite[p.180]{Jack}.  

\subsection{Mass energy equivalence}

   One of the most spectacular and profound insights gained by special relativity is that mass
is somehow equivalent to huge amounts of energy. We show here how this amazing result finds
a natural expression in terms of the basic relationships (\ref{stvelocity}) and (\ref{horizon}) 
relating inertial systems moving at a constant relative velocity. 

Let $X=ct+vt \x$ be the history of a particle moving at a constant velocity $v$ along the positive $x$ axis in $\H$.
Then its space-time velocity is given by 
  \beq V=\frac{dX}{dt}=c+v \x =c(1+\frac{v}{c}\x )  \label{movingmass} . \eeq
On the other hand, in the inertial system
$\H^\prime =\H \e{-\phi v \x}$ the particle is at rest. Differentiating $X^\prime = X \e{-\phi \x}$, we
find that 
  \beq V^\prime = \frac{d X^\prime}{dt^\prime} =\frac{dt}{dt^\prime}V \e{-\phi \x}=
      c\frac{dt}{dt^\prime}\cosh \phi (1-\frac{v^2}{c^2})=c, \label{restmass} \eeq
from which it follows that
    \[  \frac{dt}{dt^\prime} = \cosh \phi = \frac{1}{\sqrt{1-v^2/c^2}}\ge 1. \]
    
Recalling the classical
definition of the momentum of a particle of mass $m$,
 ${\bf p}=m {\bf v}$, in light of the equations (\ref{movingmass}) and
(\ref{restmass}), the appropriate definition of the {\it space-time momentum}
of the particle is $P=mc V=(mc^2+c \bp)$ where 
\beq m:= m_0 \cosh \phi =\frac{m_0}{\sqrt{1-v^2/c^2}}.  \label{relmass}  \eeq
is the {\it relative mass} of the
{\it rest mass} $m_0$ moving at the velocity of $v\x$, \cite{H74}.

   Let us calculate the work done in
accelerating a rest mass $m_0$ along some space trajector $\bx(t)$ to the speed of light $|{\bf v}| =c$ 
in some inertial system $\H$. In order to keep things as simple as possible, we choose the 
space trajectory $\bx(t)=\frac{1}{2}t^2 a_0 \x$ along the positive $x$-axis having space
velocity ${\bf v}=\frac{d \bx}{dt}=a_0t \x$, and space acceleration ${\bf a}=\frac{d^2 \bx}{dt^2}=a_0 \x$. 
Newton's 2nd law states that $F=m{\bf a}$. We will utilize this law where $m$ is the relative mass
defined in (\ref{relmass}). Thus,
   \beq Work = \int m {\bf a}\cdot d \bx =a_0  \int_0^{c^2/2a_0} m dx =
                \int_0^c m v dv = m_0 \int_0^c \frac{vdv}{\sqrt{1-v^2/c^2}}=m_0c^2 \label{massenergy}, \eeq
 as follows from the relations $v=ta_0$ and $dx=v dt=\frac{v}{a_0}dv$. This result takes
 on unexpected new significance in light of the recent work \cite{S5}.

\section{Conclusions}
   We have done all of our calculations in the space-time algebra of observables
$\C_3$. Space-time can be further simplified by considering the complex vector algebra
to be the even subalgebra of Hestenes' the higher dimensional 
{\it space-time algebra} \cite{H74}. 
Whereas calculations become simpler, only the even elements of the space-time algebra
have the direct physical meaning of observables. 
The relationships between these various space-time algebras is explored in 
 \cite{BS}.

 \section*{Acknowledgements}
 The author thanks Dr. Guillermo Romero, Academic Vice-Rector, and Dr. Reyla Navarro, Chairwomen of
   the Department of Mathematics, at the Universidad de Las Americas for continuing 
   support for this research. He and is a member of SNI 14587. 
       (URL: http://www.garretstar.com)

\end{document}